# Evaluating Metrics for Standardized Benchmarking of Remote Presence Systems

Charles Peasley, Iowa State University
Rachel Dianiska, Iowa State University
Emily Oldham, Iowa State University
Nicholas Wilson, Iowa State University
Stephen Gilbert, Iowa State University

Peggy Wu, United Technologies Research Center
Brett Israelsen, United Technologies Research Center
James Oliver, Iowa State University

**Seeking Travel Replacement Thresholds**
To reduce the need for business-related air travel and its associated energy consumption and carbon footprint, the U.S. Department of Energy's ARPA-E is supporting a research project called SCOTTIE - Systematic Communication Objectives and Telecommunications Technology Investigations and Evaluations. SCOTTIE tests virtual and augmented reality platforms in a functional comparison with face-to-face (FtF) interactions to derive travel replacement thresholds for common industrial training scenarios. The primary goal of Study 1 is to match the communication effectiveness and learning outcomes obtained from a FtF control using virtual reality (VR) training scenarios in which a local expert with physical equipment trains a remote apprentice without physical equipment immediately present. This application scenario is commonplace in industrial settings where access to expensive equipment and materials is limited and a number of apprentices must travel to a central location in order to undergo training. Supplying an empirically validated virtual training alternative constitutes a readily adoptable use-case for businesses looking to reduce time and monetary expenditures associated with travel. The technology used for three different virtual presence technologies was strategically selected for feasibility, relatively low cost, business relevance, and potential for impact through transition. The authors suggest that the results of this study might generalize to the challenge of virtual conferences.

One of the major hurdles for any remote collaboration is supporting the behavioral intricacies of natural social interactions. The selection decision of the communication medium is multifaceted (e.g. El-Shinnawy & Markus, 1997). The communication outcome can further be influenced by evolving compensatory behaviors that collaborators make to accommodate for limitations of that medium (Kock et al., 2006). The present work aims to investigate support for abstract communication objectives using current commercially available platforms. As no previous work identified a comparable set of assessable benchmarks for such platforms, a novel Communications Objective Model (COM) was developed based on over 150 publications that systematically characterizes the psychosocial constructs of effective telecommunications, how these constructs manifest in behavioral markers, and several methodological approaches supporting their measurement.

This COM research is currently under review and can be discussed at the Virtual Conferencing Workshop. SCOTTIE's Study 1, described below, is in progress and designed to explore the effectiveness of the COM metrics for evaluating remote presence systems.

**SCOTTIE Study 1: Comparing Three Systems for Remote Trainer Presence and Training Effectiveness**

Each participant will receive maintenance training on a set of circuit systems, such as an LED display, a spinning motor, photoresistor, or a potentiometer. Each trial will consist of a short training session in which an expert confederate describes the particular system and relevant troubleshooting procedures, followed by the administration of several surveys, and then a test assessing participants' ability to fix a malfunctioning circuit without assistance and under time-constraints. Critically, each training session will be conducted through a particular randomly assigned and counterbalanced communication medium. The experimental task of performing maintenance on various breadboard circuit systems was selected for its ecological realism, scalable level of complexity, spatial and procedural memory components, and ability to equate cognitive demand between tasks. A mobile app tablet interface will provide diagnostic information for the participant as they complete the task.

There will be four experimental within-subject conditions used to assess the impact of using different platforms and technologies on facilitating procedural and spatial knowledge acquisition for conducting maintenance on the circuit systems: Condition 0) FtF co-location, Condition 1) a lean teleconferencing medium, and Condition 2) a virtual environment. These conditions vary in their level of artificiality, a broad term here encompassing user agency, embodiment, and media richness (Milgram et al., 1994; Benford et al., 1998; Ens et al., 2019). Agency refers to the extent to which a participant can make modifications to his or her view of the world; embodiment refers to the visual and auditory characteristics of the human operator, where high embodiment values represent greater capability to represent gestures and body language; media richness is the extent to which the work environment is represented and includes factors such as accuracy, video resolution, and audio quality. Together, these independent variables are anticipated to impact the degree to which the technology platform supports communication metrics as described by the COM and to influence the overall user experience as a result.

*Condition 0: Face to Face*
In the FtF control condition, the trainer and apprentice (i.e., participant) will be co-located for training, whereas all other conditions involve remote communication. Agency, embodiment, and media richness levels are at the maximum possible value for the in-person training.

*Condition 1: Lean Medium Telecommunications*
In the lean medium condition, standard commercially available telecommunications tools such as Zoom or skype is used with live video stream on a standard desktop or laptop computer. The apprentice receives instruction through an audio stream and a forehead webcam perspective of the physical workspace. This condition affords the apprentice low agency due to the inability to move about the space and to actively manipulate perspective. User embodiments are medium due to visual realism but still lower than FtF. Media Richness is considered low due to the limited field of view of the the camera.

*Condition 2: VR*
During the VR training conditions, the remote apprentice will don an HTC Vive while the expert trainer will don a Microsoft HoloLens. Two viewpoints are afforded to the apprentice that can be chosen at will by the apprentice: either a static image of the workspace taken from a 360° video camera that continuously updates with an avatar representation of the expert, or a live video feed from the trainer's forehead camera. Simple avatars of the trainer and apprentice will be represented for each other in the same 3D space. Gestural information will be delivered through these avatars, and audio streams are delivered through in-house speakers and microphones. The VR condition affords medium agency because the apprentice can control point of view, embodiments are low due to low avatar realism, and media richness is considered medium since spatial information about the environment and the operator's interactions with the environment are replicated.
.

**Dependent Variables Based on the COM**
Dependent variables include electrophysiological data, qualitative measures of user experience, and objective measures of performance during testing of the trained task. Electrophysiological data will be recorded from participants using the Empatica E4 wristwatch and facial electromyography (EMG). Qualitative surveys administered following each training session will assess trust, shared situational awareness and mental models, engagement, co-presence, and mental workload or ease of use. Objective measures include task performance metrics such as time to completion, accuracy, and number of errors. Participants will also complete a delayed interval posttest in which performance on the same set of maintenance tasks is assessed again two weeks after completing the original training session. This additional component addresses a gap in the existing literature because apprentices are often challenged with applying knowledge after some period of delay following training.

The dependent variable scores for each independent training condition will be statistically compared with the baseline condition FtF, co-located training serving as the benchmark for each measure. The following discussion outlines the expected dependent variable outcomes according to the tested communication technologies. Based on previous findings in the literature: task performance is expected to vary according to levels of agency during the training, wherein higher levels of agency lead to increased knowledge retention; trust and rapport increase with avatar realism and with embodiment; co-presence and shared situational awareness benefit from increased agency, embodiment, and media richness; cognitive workload decreases with increased agency and increased environmental richness. Therefore, task performance, trust and rapport, co-presence, shared situational awareness, and ease of use should be best for VR, followed by the telecommunications conditions.

This experiment will assess the potential for virtual and augmented reality technologies to improve upon existing teleconferencing platforms in delivering equipment training to remote users and compare the respective capacities of each relative to in-person training. The most direct goal is to provide a viable and cost-efficient software and hardware alternative that reduces the need and decision for business travel under the constraint of training with immobile equipment.

For this Virtual Conferencing Workshop, the research team is prepared to present virtually, and one of the authors may be attending the conference live, which could facilitate this process. The team would look forward to demonstrating the remote presence systems and described the COM measures used to evaluate their success.

**Acknowledgments:**

This work is funded by the Department of Energy's Advanced Research Projects Agency-Energy (ARPA-E) and United Technologies Research Center (UTRC). We would like to thank Dr. Rachel Slaybaugh and Mr. Reid Rusty Heffner for their continued guidance and support.